\documentclass[conference]{IEEEtran}
\IEEEoverridecommandlockouts
\usepackage{cite}
\usepackage{amsmath,amssymb,amsfonts}
\usepackage{algorithmic}
\usepackage{graphicx}
\usepackage{textcomp}
\def\BibTeX{{\rm B\kern-.05em{\sc i\kern-.025em b}\kern-.08em
    T\kern-.1667em\lower.7ex\hbox{E}\kern-.125emX}}

\usepackage{parskip} 
\usepackage{setspace} 
\usepackage{emptypage} 
\usepackage{multicol} 
\usepackage{titlesec}
\titlespacing{\section}{0pt}{3.3ex}{2ex}
\titlespacing{\subsection}{0pt}{3.3ex}{1.65ex}
\titlespacing{\subsubsection}{0pt}{3.3ex}{1ex}
\usepackage{color}

\usepackage[english]{babel} 
\usepackage[utf8]{inputenc} 
\usepackage[T1]{fontenc} 
\usepackage[11pt]{moresize} 
\usepackage{amsfonts} 
\usepackage{blindtext}
\usepackage{changepage}

\usepackage[dvipsnames]{xcolor}
\usepackage{soul}
\usepackage[normalem]{ulem} 

\usepackage{graphicx}
\usepackage{transparent} 
\usepackage{eso-pic} 
\usepackage{tikz} 
\usetikzlibrary{}
\graphicspath{{./Images/}} 
\usepackage{graphicx,caption}
\usepackage{caption} 
\usepackage[capposition=bottom]{floatrow}
\usepackage{subcaption}
\usepackage{xcolor} 
\usepackage{amsthm,thmtools,xcolor} 
\usepackage{float}

\usepackage{amsmath}
\newtheoremstyle{italicstyle} 
  {3pt} 
  {3pt} 
  {\itshape} 
  {1.5em} 
  {\itshape} 
  {:} 
  {.5em} 
  {} 
{
\theoremstyle{italicstyle}
\newtheorem{assumption}{Assumption}
}
\usepackage{amsthm}
\usepackage{amssymb}
\usepackage{amsfonts}
\usepackage{bm}
\usepackage{bbm}
\usepackage[overload]{empheq} 
\usepackage{fix-cm} 
\usepackage{blkarray}
\usepackage{nicematrix}
\usetikzlibrary{decorations.pathreplacing,calligraphy}
\usepackage{easybmat}
\usepackage{multirow,bigdelim}
\usepackage{siunitx}
\usepackage{mathtools}

\newtheorem{prob}{Problem}
\DeclarePairedDelimiter\abs{\lvert}{\rvert}%
\DeclarePairedDelimiter\norm{\lVert}{\rVert}%
\makeatletter
\let\oldabs\abs
\def\abs{\@ifstar{\oldabs}{\oldabs*}}
\let\oldnorm\norm
\def\norm{\@ifstar{\oldnorm}{\oldnorm*}}
\makeatother
\usepackage{arydshln}

\usepackage{tikz}
\usetikzlibrary{decorations.pathreplacing}
\usetikzlibrary{fit}
\tikzset{%
highlight/.style={rectangle,rounded corners,fill=green!15,draw,
fill opacity=0.5,thick,inner sep=0pt}
}

\usepackage{array}

\usepackage{tabstackengine}
\setstackEOL{\cr}

\usepackage{tabularx}
\usepackage{longtable} 
\usepackage{colortbl}
\usepackage{subcaption}

\usepackage{algorithm}
\usepackage{algorithmic}

\usepackage[colorlinks=true,linkcolor=black,anchorcolor=black,citecolor=black,filecolor=black,menucolor=black,runcolor=black,urlcolor=black]{hyperref} 
\usepackage[square, numbers, sort&compress]{natbib} 
\usepackage{cleveref}  
\usepackage{pdfpages} 
\usepackage{afterpage}
\usepackage{lipsum} 
\usepackage{fancyhdr} 
\usepackage{indentfirst}
\usepackage{subcaption}
\begin{document}

\title{Intermodal Network of Autonomous Mobility-on-Demand and Micromobility Systems\\
}

\author{\IEEEauthorblockN{1\textsuperscript{st} Seyyed Jalaladdin Abbasi Koumleh}
\IEEEauthorblockA{\textit{department of mechanical engineering} \\
\textit{Politecnico di Milano}\\
Milan, IT \\
seyyedjalaladdin.abbasi@mail.polimi.it}
\and
\IEEEauthorblockN{2\textsuperscript{nd} Fabio Paparella}
\IEEEauthorblockA{\textit{department of mechanical engineering} \\
\textit{Eindhoven University of Technology}\\
Eindhoven, ND \\
f.paparella@tue.nl}

}

\maketitle

\begin{abstract}
This paper studies models for Autonomous Micromobility-on-Demand (AMoD), a paradigm in which a fleet of autonomous vehicles delivers mobility services on demand in conjunction with micromobility systems. Specifically, we introduce a network flow model to encapsulate the interaction between AMoD and micromobility under an intermodal connection scenario. The primary objective is to analyze the system's behavior, optimizing passenger travel time. Following this theoretical development, we apply these models to the transportation networks of Sioux Falls, enabling a quantifiable evaluation of the reciprocal influences between the two transportation modes. We found that increasing the number of vehicles in any of these two modes of transportation also incentivizes users to use the other. Moreover, increasing the rebalancing capacity of the micromobility system will make the AMoD system need less rebalancing.
\end{abstract}


\section{Introduction}

\begin{figure}[!tb]
\centering
{\includegraphics[width=\linewidth]{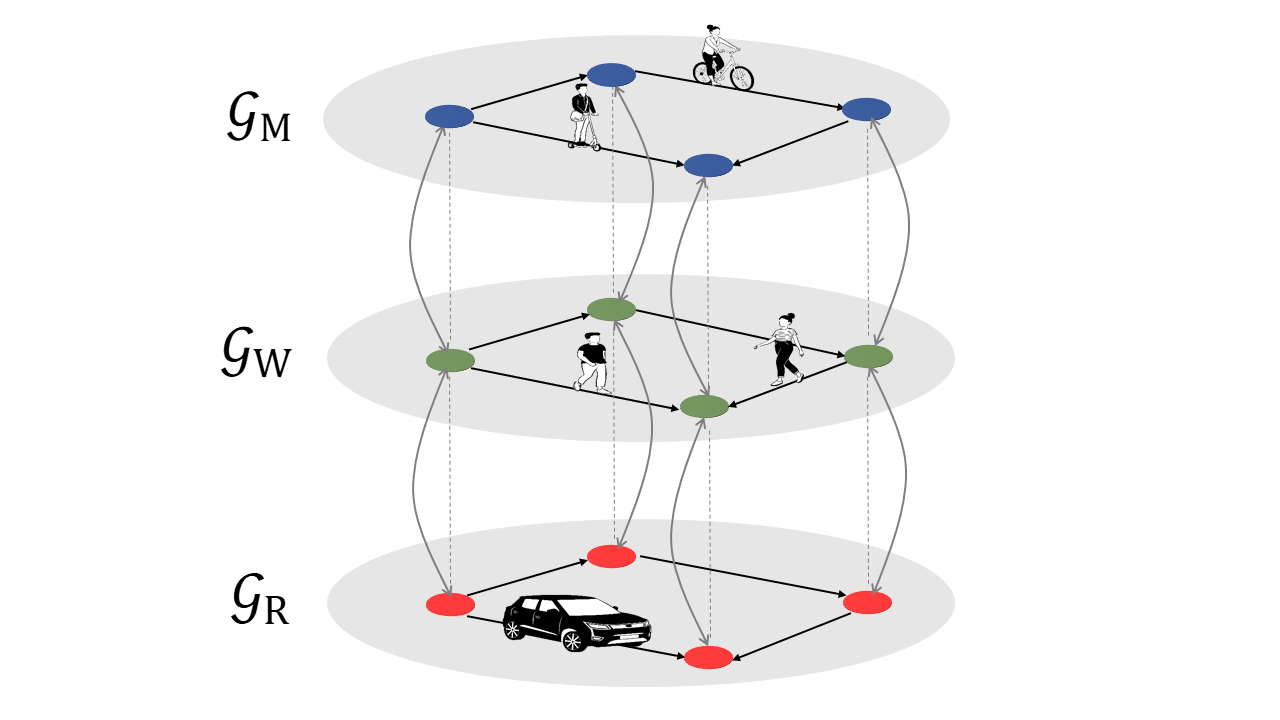}}%
\caption[Intermodal network representation]{Intermodal network representation: the intermodal network consists of walking, micromobility, and AMoD graphs. The colored dots denote intersections or stops, and the black arrows represent road links or pedestrian pathways. The dotted lines denote geographically close nodes, while the gray curved arrows are the mode-switching arcs connecting them.\protect}
\label{fig: intermodal and multimodal networks representation}%
\end{figure}

Urban transit is on the brink of a transformative shift with the introduction of Autonomous Mobility on Demand (AMoD). This innovative system, managed by central command centers, optimizes routes to transport passengers across cities efficiently. The critical advantage of AMoD is its potential to significantly reduce travel times and alleviate congestion, offering a superior alternative to traditional ride-sharing and taxi services. By allowing for direct route control and autonomous rebalancing, which is the autonomous relocation of the vehicle after serving one travel to reach the origin of the next one, it eliminates the need for driver repositioning, maximizing Vehicle utilization rates and enhancing operational efficiency and flexibility. Despite its promise, the standalone operation of AMoD systems might inadvertently contribute to traffic congestion through shifts in transport modal preferences \cite{bi1,bla12}. To harness its full potential for sustainable urban mobility, integrating AMoD with existing transit systems is essential, ensuring it complements rather than supplants these systems to create a cohesive, congestion-free travel ecosystem. Integrating micromobility systems, such as electric scooters, bikes, and e-bikes, is particularly advantageous in this context. Micromobility offers a sustainable solution for short-distance travel, especially in densely populated urban areas. These lightweight vehicles reduce the number of cars on the road, lessening congestion and pollution. By manually rebalancing the micromobility vehicles based on real-time demand and ensuring that vehicles are available where and when needed, we can further reduce the dependency on personal vehicles. This will contribute significantly to urban air quality and carbon footprint reduction. This study explores the potential of integrating Autonomous Mobility-on-Demand (AMoD) and micromobility systems to minimize passengers' travel time. By combining these two modes of transportation into an intermodal network in which the user can freely transition between all existing forms of transportation, this research aims to extend the understanding of network flow models with multi-commodity flows, highlighting the interactions and possible benefits of integrating AMoD and micromobility systems (Figure \ref{fig: intermodal and multimodal networks representation}).

\textit{Related Literature: }Multi-commodity network flow models \cite{bla1, bla2, bla3} are widely utilized for characterizing and managing transportation systems, especially within the dynamic Autonomous Mobility-on-Demand (AMoD) sector. These models are particularly advantageous over queuing-theoretical models \cite{bla4, bla5, bla6} and simulation-based models \cite{bla7, bla8, bla9} because they can incorporate a diverse array of constraints and are compatible with commercial optimization solvers. These models have been widely used in problems ranging from control of AMoD systems in road networks \cite{bi1, bla1,bla2,bla3, bla10} to smart charging \cite{bla11}, ride-pooling assignment problems \cite{bla14}, and accessibility fairness \cite{bla13}.
Several studies on micromobility network flows have been conducted in isolation. In \cite{blm1}, the authors determined the optimal location of bike stations by modeling and solving a multi-objective optimization problem. In contrast, \cite{blm2} includes a similar study on a smaller scale, incorporating the geographic information system (GIS) and level of traffic stress (LTS) ratings. The authors of \cite{blm3} have proposed the shortest time and distance routes while keeping the user away from potential COVID-19 transmission spots. Research on intermodal and multimodal AMoD transportation networks focusing on multi-commodity network flow models still needs to be more extensive. The paper \cite{blj1} describes a network flow model for jointly optimizing the AMoD routing and rebalancing strategies in a congestion-aware fashion. The authors of \cite{blj2} have proposed a network flow model to illustrate the system behavior when AMoD vehicles and public transport provide service to users in an intermodal conjunction. They then showed that the intermodal AMoD configuration has several benefits over using AMoD in isolation. In \cite{blj3}, the authors have completed the previous work by studying the vehicles’ size and powertrain impact on system efficiency and routing decisions. Zgraggen, Tsao, Salazar, Schiffer, and M. Pavone, in \cite{blj4}, use a model predictive control (MPC) algorithm to route customers and vehicles through the network to minimize customers' travel time on a model that captures the operation of an AMoD system in coordination with public transit. To the authors’ knowledge, no study has been performed to capture the interaction of AMoD and micromobility vehicles.

\textit{Statement of Contribution: }This work contributes twofold: first, we present a multicommodity network flow optimization model that casts the joint operations of AMoD and micromobilty systems to minimize the customer's travel time. Then, we present a case study for Sioux Falls, accounting for the impact of the micromobility network and AMoD vehicles on travel times. 

\textit{Organization: }The remainder of this paper is organized as follows: Section II presents the construction of the multi-commodity flow model for the AMoD-Micromobilty-Walking network using the elements of graph theory and then casts the optimization problem accordingly. In Section III, we implement our approach in Sioux Falls' case studies. In section III, we outline the significant results of this work.

\section{NETWORK FLOW MODEL}
In this section, first, we introduce the network flow model presented in \cite{blj3} and extend it to include the micromobilty system. Then, we discuss the representation of travel time. Subsequently, we reformulate the problem to capture the joint interaction of AMoD with micromobility. Lastly, we briefly discuss the model.
\subsection{Multi-commodity Flow Model}
Consider a transportation system with three modes: walking, micromobility, and AMoD. We denote $\mathcal{G}_\mathrm{W} = (\mathcal{V}_\mathrm{W},\mathcal{A}_\mathrm{W})$ as the walking network, $\mathcal{G}_\mathrm{M} = (\mathcal{V}_\mathrm{M},\mathcal{A}_\mathrm{M})$ as the micromobility network and $\mathcal{G}_\mathrm{R} = (\mathcal{V}_\mathrm{R},\mathcal{A}_\mathrm{R})$ as the road network, where $(\mathcal{V}_\mathrm{W},\mathcal{A}_\mathrm{W})$, $(\mathcal{V}_\mathrm{M},\mathcal{A}_\mathrm{M})$ and $(\mathcal{V}_\mathrm{R},\mathcal{A}_\mathrm{R})$ are the sets of vertices and arcs on the walking, micromobility, and road networks, respectively. Moreover, we represent by $N_\mathrm{W}$ and $E_\mathrm{W}$, the number of nodes and the number of edges of the walking graph, i.e., $N_\mathrm{W} = \abs{\mathcal{V}_\mathrm{W}}$ and $E_\mathrm{W} = \abs{\mathcal{A}_\mathrm{W}}$. Similarly, $N_\mathrm{M}$, $N_\mathrm{R}$, and  $E_\mathrm{M}$, $E_\mathrm{R}$ denote the number of nodes and the number of edges of micromobility and road networks. To define the corresponding incidence matrices, we denote by $B_\mathrm{W} \in \{-1,0,1\}^{\abs{\mathcal{V}_\mathrm{W}} \times \abs{\mathcal{A}_\mathrm{W}}}$, $B_\mathrm{M} \in \{-1,0,1\}^{\abs{\mathcal{V}_\mathrm{M}} \times \abs{\mathcal{A}_\mathrm{M}}}$ and $B_\mathrm{R} \in \{-1,0,1\}^{\abs{\mathcal{V}_\mathrm{R}} \times \abs{\mathcal{A}_\mathrm{R}}}$, the incidence matrices of walking, micromobility and road networks.

We denote by $\mathcal{A}_\mathrm{S}$ the set of switching arcs where  $\mathcal{A}_\mathrm{S} = \mathcal{A}_\mathrm{WM} \cup  \mathcal{A}_\mathrm{WR}$, in which $\mathcal{A}_\mathrm{WM} \subseteq  \left(\mathcal{V}_\mathrm{W} \times \mathcal{V}_\mathrm{M}\right) \, \cup \,  \left(\mathcal{V}_\mathrm{M} \times \mathcal{V}_\mathrm{W}\right)$ and $\mathcal{A}_\mathrm{WR} \subseteq \left(\mathcal{V}_\mathrm{W} \times \mathcal{V}_\mathrm{R}\right)\, \cup \, \left(\mathcal{V}_\mathrm{R} \times \mathcal{V}_\mathrm{W}\right)$ represent switching arcs between walking-micromobility and walking-road. We represent by $N_\mathrm{S}$ and $E_\mathrm{S}$, the total number of switching nodes and edges, and by $B_\mathrm{S}\in \{-1,0,1\}^{\abs{\mathcal{V}} \times \abs{\mathcal{A}_\mathrm{s}}}$, the incidence matrix of the switching network.

We also define the supergraph $\mathcal{G} = (\mathcal{V},\mathcal{A})$, which includes all the vertices and arcs of our network. In other words, G will be the union of all sets and arcs, thus $\mathcal{V} = \mathcal{V}_\mathrm{W}  \cup  \mathcal{V}_\mathrm{M}  \cup  \mathcal{V}_\mathrm{R} $ and $\mathcal{A} = \mathcal{A}_\mathrm{W}  \cup  \mathcal{A}_\mathrm{M}  \cup  \mathcal{A}_\mathrm{R}  \cup  \mathcal{A}_\mathrm{S}$. We denote by $N$ and $E$, the total number of nodes and edges, and by $\mathrm{B} \in \{-1,0,1\}^{\abs{\mathcal{V}} \times \abs{\mathcal{A}}}$ the incidence matrix of the supergraph $\mathcal{G}$. Considering the structural properties of walking, micromobility, and road networks in urban environments, we make the following assumption:
\begin{assumption}\label{as:1}
The graphs $\mathcal{G}$, $\mathcal{G}_\mathrm{W}$, $\mathcal{G}_\mathrm{M}$ and $\mathcal{G}_\mathrm{R}$ are strongly connected.
\end{assumption}

Considering the mesoscopic nature of our study, we define the request $r$ as the triple $(\mathrm{o},\mathrm{d}, \alpha) \in \mathcal{V} \times \mathcal{V} \times \mathbb{R}_+$ in which $\mathrm{o} \in \mathcal{V}$ and $\mathrm{d} \in \mathcal{V}$ are respectively the origin and destination nodes and $\alpha \in \mathbb{R}_+$ is the request rate defined as the number of requests for travel from node $\mathrm{o}$ to node $\mathrm{d}$ per unit of time. We collect all $M$ demands in a demand set $\mathcal{R}$, such that: $\mathcal{R} = \{\mathrm{r}_m : \mathrm{r}_m = (\mathrm{o}_m,\mathrm{d}_{m}, \alpha_m), \; m \in \mathcal{M}=\{1,...,M\}\}$. We make the following assumption:

\begin{assumption}\label{as:2}
All requests appear on the walking digraph, i.e. $\mathrm{o}_m,\mathrm{d}_m \in \mathcal{V}_\mathrm{W}, \, \forall m \in M$.
\end{assumption}

We denote by $x_{ij}^{\mathrm{m}}$ the user flow, which is defined as the number of users passing through the edge $(i,j) \in \mathcal{A}$ per unit time per travel request $\mathrm{r}_m$. For the sake of compactness of the mathematical notations and expressions, we define $X_{ij}$ as: 

\begin{equation}
\begin{aligned}
X_{ij}^{} 
=
\sum_{m \in \mathcal{M}} x_{ij}^{m} 
\qquad
\forall (i,j) \in \mathcal{A}
\end{aligned}
\end{equation}

Moreover, We define $\beta_{i}^{\mathrm{in}}$ and $\beta_{i}^{\mathrm{out}}$ as the number of the micromobility rebalancing vehicles fed to and withdrawn from the node $i \in \mathcal{V}_\mathrm{M}$ per unit time by the operator respectively. Similarly, we denote $x_{ij}^0$ as the rebalancing flow of the AMoD vehicles for the edge $(i,j) \in \mathcal{A}_\mathrm{R}$. We consider $h_{\mathrm{S},ij}$ for $(i,j) \in \mathcal{A}_\mathrm{S}$ as the switching capacity constraint. By $\beta_\mathrm{M}$, we denote the number of micromobility vehicles the operator can rebalance per unit of time in each node, and $\beta_\mathrm{M}^\mathrm{tot}$ is considered as the total rebalancing vehicles available in the micromobilty system. We also define $n_\mathrm{M}$ and $n_\mathrm{R}$ as the total number of vehicles in the micromobility and road networks. 

\subsection{Travel Time}
We denote by $t_{ij}^0: \mathbb{R}_+^{|\mathcal{A}|} \mapsto \mathbb{R}_+$ the free-flow time which is the time it takes to travel through a given arc freely in the absence of indigenous and exogenous traffic. If $l_{ij}$ is the length of the edge $(i,j) \in \mathcal{A}$, and $v_{max}$ is the travel speed on that edge, then: 

\begin{equation}\label{eq: nominal travel time}
t_{ij}^\mathrm{0} = \frac{l_{ij}}{v_{max}} \qquad \forall  (i,j) \in \mathcal{A}
\end{equation}

 Moreover, the travel time $t_{ij}: \mathbb{R}_+^{|\mathcal{A}|} \mapsto \mathbb{R}_+$ is the actual time it takes for the vehicle to pass through the edge $(i,j) \in \mathcal{A}$. We assume there is no congestion except in the road network; therefore: 

 \begin{equation}\label{eq: equality of travel time}
t_{ij} = t_{ij}^\mathrm{0} \qquad \forall  (i,j) \in \mathcal{A}_\mathrm{W}  \cup \mathcal{A}_\mathrm{M}
\end{equation}

We assume no private vehicles are in our network to model the travel time on the road network. If $X_{ij}^\mathrm{road}$ is the total vehicle flow on the edge $(i,j) \in \mathcal{A}_\mathrm{R}$, then:

\begin{equation}\label{eq: total road flow}
X_{ij}^\mathrm{road} = X_{ij} + x_{ij}^\mathrm{0}  \qquad \forall  (i,j) \in \mathcal{A}_\mathrm{R}
\end{equation} 

Therefore, to adjust the travel time to account for free-flow speed and road congestion, one can write:

\begin{equation}\label{eq: actual time travel time}
t_{ij}(X_{ij}^\mathrm{road}) = t_{ij}^0 \cdot f(X_{ij}^\mathrm{road}) \qquad \forall  (i,j) \in \mathcal{A}_\mathrm{R}
\end{equation} 

Where $f(.)$ is a volume delay function and is positive, strictly growing, and continuously differentiable. A widely used delay function by urban planners and researchers is the approach for the Bureau of Public Roads (BPR) \cite{bm1} in which $t_{ij}(X_{ij}^\mathrm{road}) = t_{ij}^0(1 + 0.15(X_{ij}^\mathrm{road}/h_{\mathrm{R},ij})^4)$, where $h_{\mathrm{R},ij}$ is the link capacity. However, to have a convex problem, we use a piece-wise approximation of this function proposed in \cite{bm2}.

\subsection{Objective and Optimization Problem}
We present the I-AMoD optimization problem as follows:

\begin{prob}\label{prob: problem 1}
(Intermodal Optimization Problem) : Given the set of transportation demands $\mathcal{R}$, the optimal user, vehicles, and rebalancing flows result from the optimization problem
\begin{subequations}\label{eq: Optimization General}
\begin{equation}\label{eq: Objective Function}
\begin{aligned}
&\min_{x_{ij}^{m}, \beta_{i}^{\mathrm{in}}, \beta_{i}^{\mathrm{out}}, x_{ij}^{0}}
\quad
\sum_{(i,j)\in\mathcal{A}} t_{ij}\cdot X_{ij}
\end{aligned}
\end{equation}

\begin{flushleft}
s.t. 
\end{flushleft}
\vspace*{-\baselineskip}   

\begin{flalign} \label{b :eq: flow conserv}
& \qquad \sum_{i:(i,j) \in \mathcal{A}} x_{ij}^{m} + \mathbbm{1}_{j=\mathrm{o}_m} \cdot \alpha_m = \sum_{k:(j,k) \in \mathcal{A}} x_{jk}^{m} + \mathbbm{1}_{j=\mathrm{d}_m} \cdot \alpha_m, &\nonumber \\
&\hspace{15em} \forall m \in \mathcal{M}, \forall j \in \mathcal{V} &
\end{flalign}

\begin{equation}\label{k :eq: Road Vehicle Balance}
\begin{split}
&\sum_{i:(i,j)\in \mathcal{A}_\mathrm{R}} \left( X_{ij}^{\mathrm{c}} + x_{ij}^0\right) = \sum_{k:(j,k)\in\mathcal{A}_\mathrm{R}}   \left( X_{jk}^\mathrm{c}+ x_{jk}^0\right), \\
&\hspace{10.5em} \forall j \in \mathcal{V}_\mathrm{R}
\end{split}
\end{equation}

\begin{equation}\label{d :eq: MM Vehicle Balance}
\begin{aligned}
\sum_{i:(i,j)\in \mathcal{A}_\mathrm{M}}
X_{ij}^{\mathrm{b}}
+
\mathbbm{1}_{j} \cdot \beta_{j}^{\mathrm{in}}
&= 
\sum_{k:(j,k)\in \mathcal{A}_\mathrm{M}}
X_{jk}^\mathrm{b} 
+
\mathbbm{1}_{j} \cdot \beta_{j}^{\mathrm{out}}, \\
&\qquad \forall j \in \mathcal{V}_\mathrm{M}
\end{aligned}
\end{equation}

\begin{equation}\label{e :eq: micromobility Rebalancing }
\begin{aligned}
\sum_{j \in \mathcal{V}_\mathrm{M}} \beta_{j}^{\mathrm{in}}
=
\sum_{j \in \mathcal{V}_\mathrm{M}} \beta_{j}^{\mathrm{out}}
\leq
\beta_\mathrm{M}^\mathrm{tot}
\end{aligned}
\end{equation}

\begin{equation}\label{m :eq: road Fleet size constraint}
\begin{aligned}
\sum_{(i,j) \in \mathcal{A}_\mathrm{R}} t_{ij} 
\cdot
X_{ij}^\mathrm{c}
\leq 
n_{\mathrm{R}}
\end{aligned}
\end{equation}

\begin{equation}\label{h :eq: MM Fleet size constraint}
\begin{aligned}
\sum_{(i,j)\in \mathcal{A}_\mathrm{M}} t_{ij}
\cdot 
X_{ij}^{\mathrm{b}}
\leq
n_{\mathrm{M}} 
\end{aligned}
\end{equation}

\begin{equation}\label{l :eq: road vehicle capacity constraint}
\begin{aligned}
X_{ij}^{\mathrm{c}} + x_{ij}^0 
\leq 
h_{\mathrm{R},ij},
\qquad
\forall (i,j) \in \mathcal{A}_\mathrm{R}
\end{aligned}
\end{equation}

\begin{equation}\label{q :eq: Switching arcs flow Constraint}
\begin{aligned}
X_{ij}^{} 
\leq
h_{\mathrm{S},ij}, 
\qquad 
\forall (i,j) \in \mathcal{A}_\mathrm{S}
\end{aligned}
\end{equation}

\begin{equation}\label{f1 :eq: beta_non_negativity}
\begin{aligned}
\beta_{j}^{k} \leq \beta_{\mathrm{M}}, \qquad \forall j \in \mathcal{V}_{\mathrm{M}}, \, k \in \{\mathrm{in}, \mathrm{out}\}
\end{aligned}
\end{equation}

\begin{equation}\label{c :eq: Xu > 0}
\begin{aligned}
x_{ij}^{m} \geq 0, \qquad \forall m\in \mathcal{M},\; \forall (i,j) \in \mathcal{A}
\end{aligned}
\end{equation}

\begin{equation}\label{o :eq: Xc > 0}
\begin{aligned}
x_{ij}^{\mathrm{c},m} \geq 0, \qquad \forall m\in \mathcal{M},\; \forall (i,j) \in \mathcal{A}_\mathrm{R}
\end{aligned}
\end{equation}

\begin{equation}\label{p :eq: X0 > 0}
\begin{aligned}
x_{ij}^0 \geq 0, \qquad \forall (i,j) \in \mathcal{A}_\mathrm{R}
\end{aligned}
\end{equation}

\begin{equation}\label{j :eq: Xs > 0}
\begin{aligned}
x_{ij}^{\mathrm{b},m} \geq 0, \qquad \forall m\in \mathcal{M},\; \forall (i,j) \in \mathcal{A}_\mathrm{M}
\end{aligned}
\end{equation}

\begin{equation}\label{g :eq: beta_non_negativity}
\begin{aligned}
\beta_{j}^{k} \geq 0, \qquad \forall j \in \mathcal{V}_{\mathrm{M}}, \, k \in \{\mathrm{in}, \mathrm{out}\}
\end{aligned}
\end{equation}

\begin{equation}\label{n :eq: Xu = Xc}
\begin{aligned}
x_{ij}^{m} = x_{ij}^{\mathrm{c},m}, \qquad \forall m\in \mathcal{M},\; \forall (i,j) \in \mathcal{A}_\mathrm{R}
\end{aligned}
\end{equation}

\begin{equation}\label{i :eq: Xu = Xs}
\begin{aligned}
x_{ij}^{m}
= 
x_{ij}^{\mathrm{b},m},
\qquad 
\forall m\in \mathcal{M},
\;
\forall (i,j) \in \mathcal{A}_\mathrm{M}
\end{aligned}
\end{equation}

\end{subequations}
\end{prob}

Constraint \textsl{(\ref{b :eq: flow conserv})} is to ensure flow conservation with demand compliance in our system, mandating that all user flows entering a given node must equal the sum of user flows exiting the node. Additionally, two other conditional terms take care of scenarios when the node is the origin or the destination of a given demand. The indicator function $\mathbbm{1}$ in these two terms is equal to 1 when the equality in the subscript holds and 0 otherwise. The conservation of AMoD and micromobility vehicles entering and exiting a node is stated by \textsl{(\ref{k :eq: Road Vehicle Balance})} and \textsl{(\ref{d :eq: MM Vehicle Balance})} respectively. Similarly, ensuring that the number of micromobilty rebalancing units entering and exiting a node is conserved and bounded is represented in \textsl{(\ref{e :eq: micromobility Rebalancing })}. Constraints \textsl{(\ref{m :eq: road Fleet size constraint})} and \textsl{(\ref{h :eq: MM Fleet size constraint})} ensure that the total number of vehicles does not exceed the predefined total. Constraint \textsl{(\ref{q :eq: Switching arcs flow Constraint})} addresses the threshold congestion, and \textsl{(\ref{l :eq: road vehicle capacity constraint})} ensures that the sum of all flows in the switching edges network remains below a capacity constraint. Constraint \textsl{(\ref{f1 :eq: beta_non_negativity})} captures the rebalancing capacity for the micromobility system in each node. Ensuring the non-negativity of flows is the purpose of \textsl{(\ref{c :eq: Xu > 0})} to \textsl{(\ref{g :eq: beta_non_negativity})}. Finally, users are assigned to AMoD and micromobility vehicles by \textsl{(\ref{n :eq: Xu = Xc})} and \textsl{(\ref{i :eq: Xu = Xs})}.

\subsection{Discussion}
A few comments are in order. The problem \ref{prob: problem 1} is considered time-invariant as the demands change slowly compared to the average total travel time of the individual trips, as in the dense urban environments \cite{basump1}. Moreover, aside from the effect of exogenous traffic on AMoD vehicles, we assume that different modes of transportation do not interfere. This is valid because we consider exclusive lanes for each mode of transportation. Furthermore, demands for origin-destination pairs are placed on the walking graph. Ride-sharing activities are also not considered; in other words, each vehicle can contain only one user. In addition, the presence of selfish users is neglected, and we assume users can be incentivized to follow a centrally controlled path to minimize average travel time. Travel time per arc is known in advance, and it takes one minute for a user to get off or ride on a vehicle. Finally, users remain in their vehicles until they reach a switching node or the destination, and passengers can only pick up or ride off vehicles in intersections, i.e., nodes.

\section{Case Study}
This section presents the implementation of our proposed framework in a case study of Sioux Falls. The data needed to construct the graph are taken from the Transportation Networks for Research repository \cite{br1}, while the travel request rates are obtained from \cite{br2}. Our model assumes that the free flow travel speed values in walking, micromobility, and road networks are 3, 15, and \qty{45}{\km\per\hour}, respectively. These values are based on careful consideration of real-world conditions. Finally, we define the users’ average travel time as follows:
\begin{equation}\label{eq: average travel time}
\begin{aligned}
t_\mathrm{avg} = \frac{ \sum_{(i,j) \in \mathcal{A}}t_{ij}X_{ij}^u}{\sum_{m \in \mathcal{M}} \alpha_m}
\end{aligned}
\end{equation}

\subsection{Effects of AMoD Fleet Size}
In this section, we present our results on the effects of the number of AMoD vehicles on the user's modal share and average travel time.

\begin{figure}[!tb]
\centering
\includegraphics[width=1\linewidth]{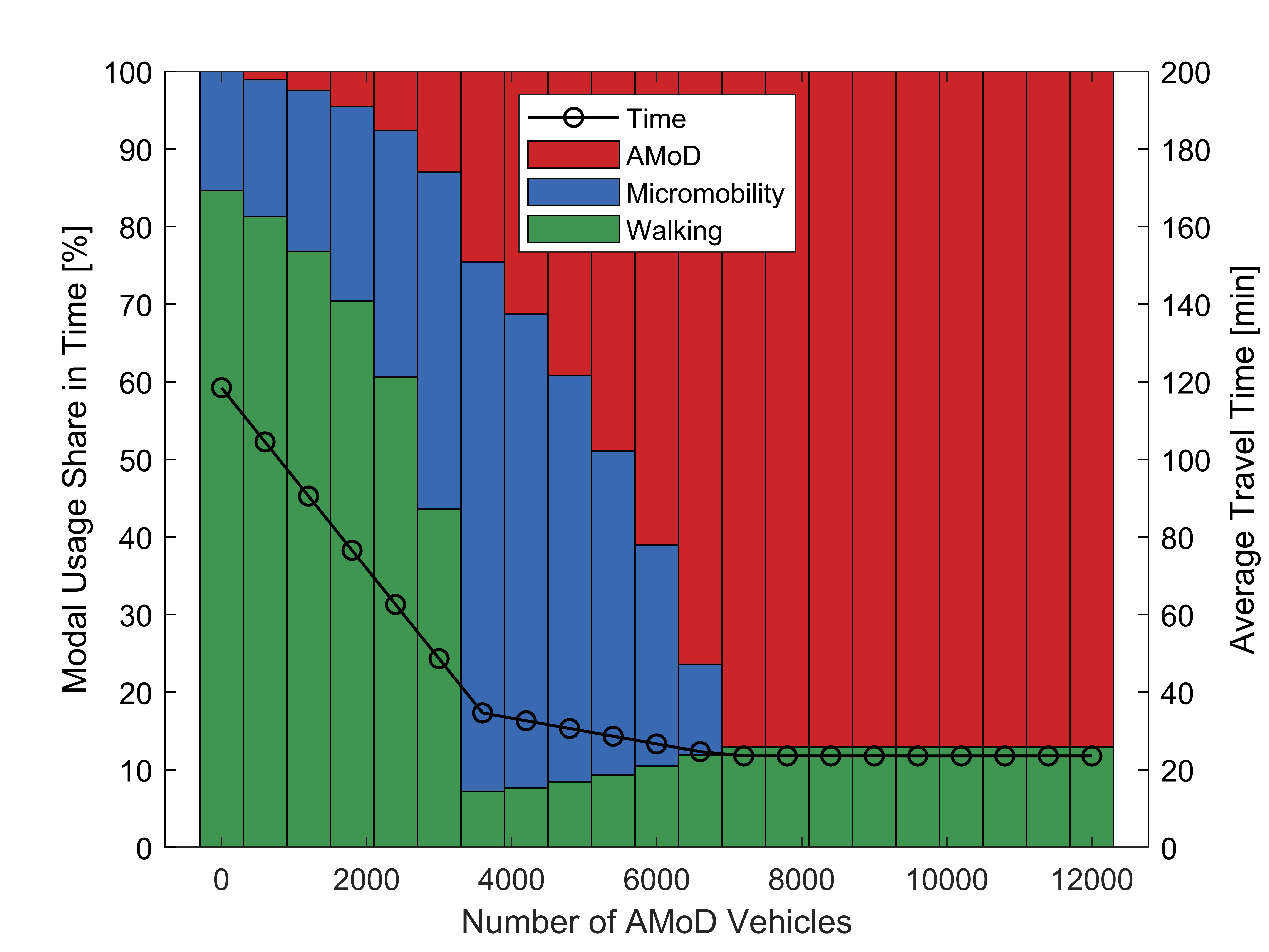} 
\caption[Sioux Falls: Time-based modal share and average travel time changing AMoD fleet size]{Time-based modal share and average travel time varying the number of AMoD vehicles. Users' modal share and average travel time converge to a steady-state value after $n_\mathrm{R} = 7000$. \protect}
\label{fig: AMoD - timeshare percentage for Sioux Falls}
\end{figure}

As depicted in Figure \ref{fig: AMoD - timeshare percentage for Sioux Falls}, initially, by increasing the number of AMoD vehicles, we have an increase in the usage share of AMoD and Micromobility systems and a decrease in the number of people walking. However, after around $n_\mathrm{R} = 4000$, we observe an increase in the usage share of walking and a decrease in the usage share of micromobility. This is because the user flow in switching arcs saturates due to the switching capacity constraint. Moreover, we reach a steady state value when $n_\mathrm{R} = 7000$, from which further increasing of AMoD vehicles does not affect the usage shares in the networks. Finally, the average travel time trend constantly decreases until it reaches a steady state value.

\subsection{Effects of micromobility Fleet Size}
\begin{figure}[!tb]
\centering
\includegraphics[width=1\linewidth]{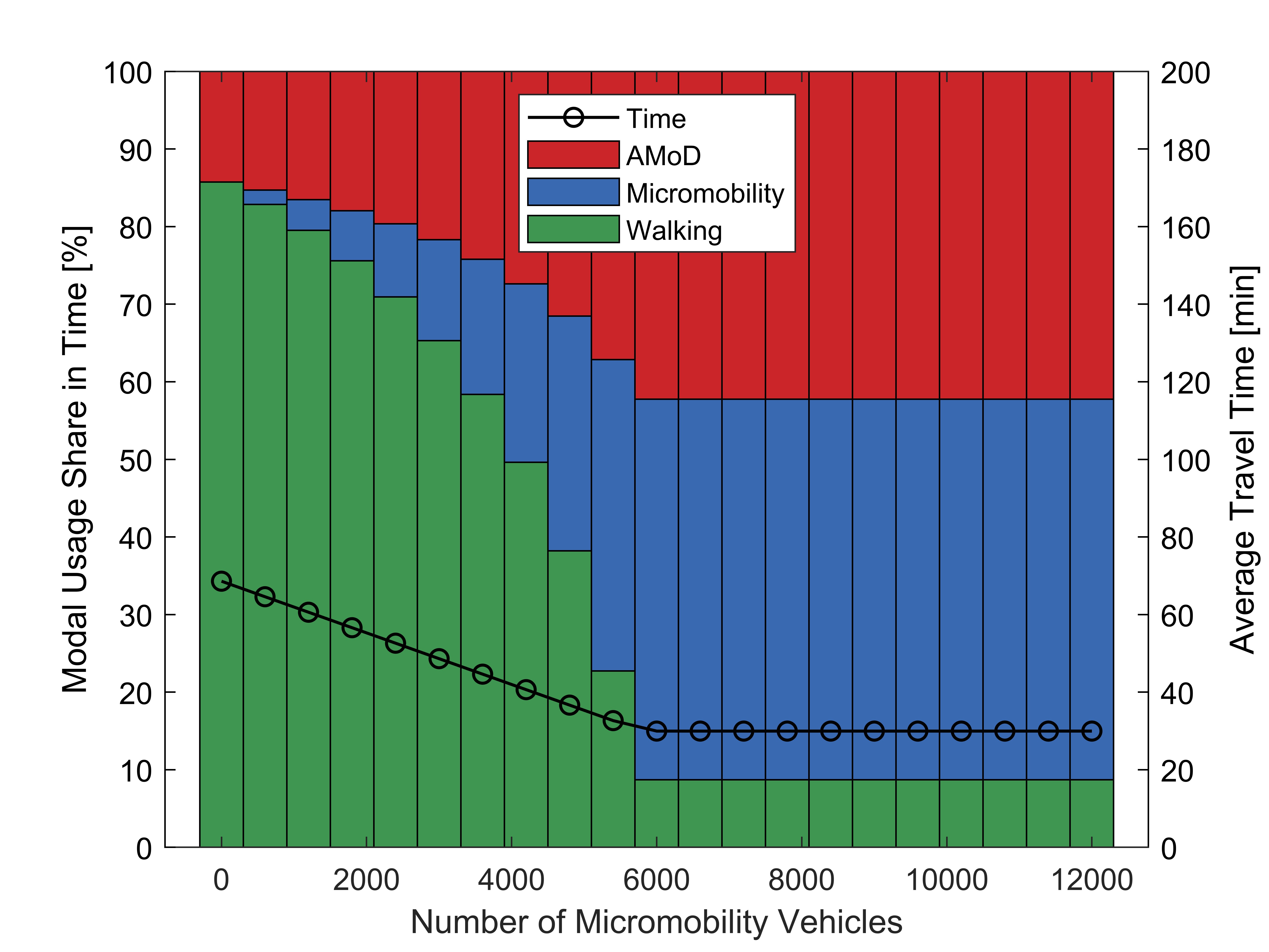} 
\caption[Sioux Falls: Time-based modal share and average travel time changing micromobility fleet size]{Time-based modal share and average travel time trend changing the micromobility fleet size. The system behavior changes, changing the AMoD to the micromobility fleet size ratio. \protect}
\label{fig: micromobility - timeshare percentage for Sioux Falls}
\end{figure}

Increasing the number of micromobility vehicles, we observe in Figure \ref{fig: micromobility - timeshare percentage for Sioux Falls} that not only improves the modal share of the micromobility network but also increases the share of the AMoD network at the expense of decreasing the total timeshare of people walking. Also, in this case, users' average travel time decreases linearly until it reaches a steady state value at $n_\mathrm{M} = 6000$, the value at which the users' share of time reaches the steady state. Regarding the average travel time, we slightly decrease the plot when we have small values for the number of vehicles, which keeps a uniform value from then on.

\subsection{Effects of Micromobility Rebalancing Capacity}
\begin{figure*}[!tb]  
\centering
    \begin{subfigure}[b]{.45\linewidth}
        \centering
        \includegraphics[width=\linewidth]{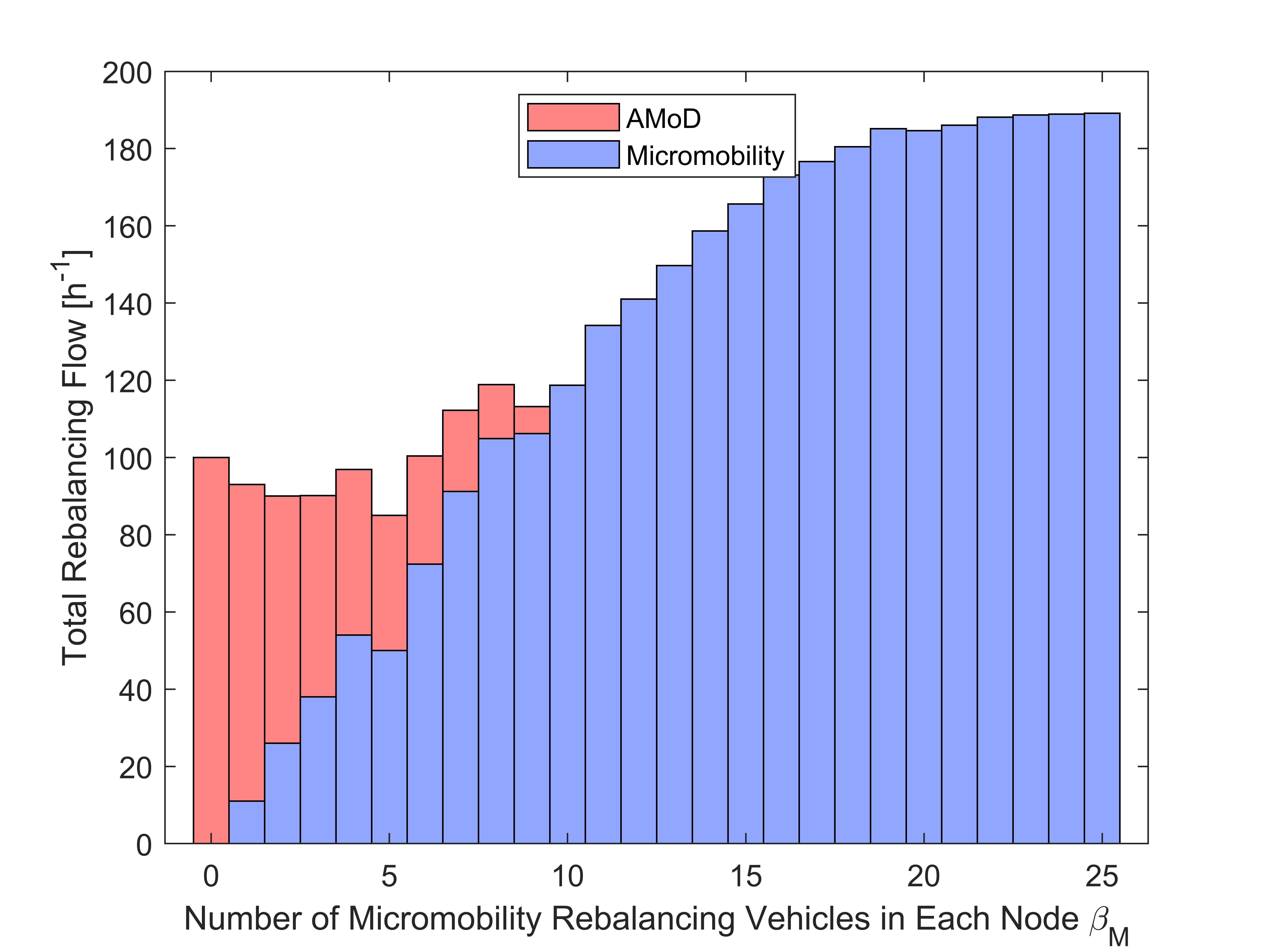}
        \caption{AMoD and micromobility total rebalancing flow change by increasing the number of each node's available micromobility rebalancing vehicles.}
        \label{fig:subfiga}
    \end{subfigure}%
    \hfill  
    \begin{subfigure}[b]{.45\linewidth}
        \centering
        \includegraphics[width=\linewidth]{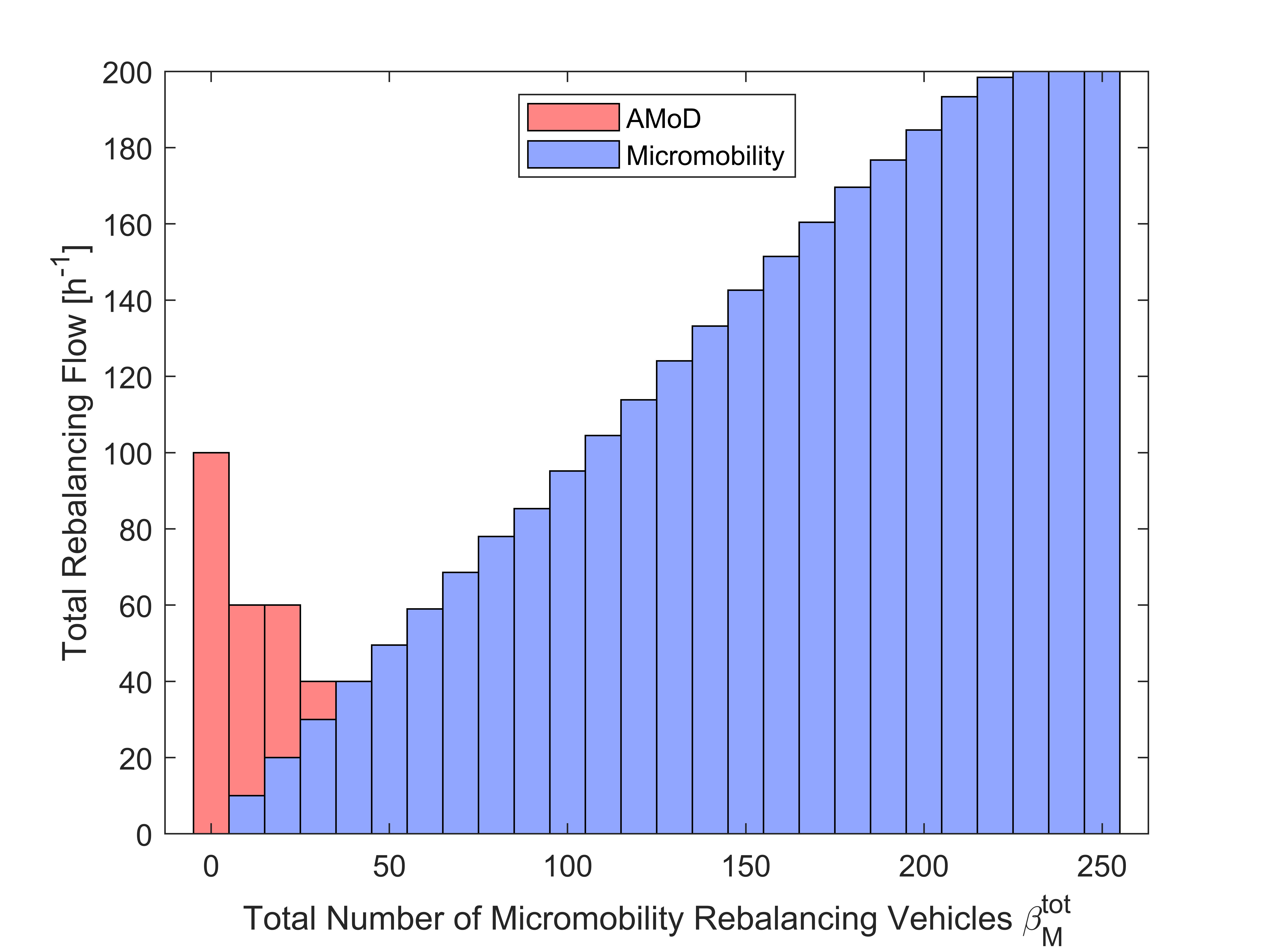}
        \caption{AMoD and micromobility total rebalancing flow change by increasing the total number of micromobility rebalancing vehicles available in the system.}
        \label{fig:subfigb}
    \end{subfigure}
\caption[AMoD and micromobility total rebalancing flow change by increasing the micromobility system's rebalancing capacity.]{AMoD and micromobility total rebalancing flow change by increasing the micromobility system's rebalancing capacity. Increasing the micromobility system's rebalancing capacity will make the AMoD vehicles rebalance less frequently. \protect}
\label{fig: results: rebalancing for changing beta_m abd beta_m^tot}
\end{figure*}

In this study, we manipulated the number of micromobility rebalancing vehicles while keeping other parameters constant. In particular, we studied the change of AMoD and micromobility rebalancing flows, and interestingly, we observed that by increasing the rebalancing capacity of micromobility, we can decrease the total rebalancing flow of the AMoD system, as shown in Figure \ref{fig: results: rebalancing for changing beta_m abd beta_m^tot}.


\section{Conclusion}
This paper studied the intermodal network of AMoD and micromobility systems. We conclude that the system's general performance improves by integrating these modes of transportation. In particular, increasing the number of AMoD vehicles incentivizes users to use the AMoD and micromobility systems until the system saturates due to the lack of parking spaces. Similarly, increasing the micromobility fleet size motivates people to use both micromobility and AMoD. Finally, by expanding the rebalancing capacity of the micromobility system, the need for AMoD rebalancing decreases, making the users' share of different modes of transportation unchanged. 

This work opens up exciting avenues for further research. For instance, economic evaluations could be integrated to assess the social costs of this joint operation. The potential benefits of ride-pooling and the impact on energy consumption factors could be explored.

\section{Acknowledgement}
The authors would like to thank Prof. Francesco Braghin for proofreading this manuscript. This research is derived from a master's thesis conducted at Politecnico di Milano in collaboration with Technische Universiteit Eindhoven.

\vspace{12pt}

\end{document}